\documentclass[runningheads]{llncs}
\usepackage[T1]{fontenc}
\usepackage{graphicx} 
\usepackage{colortbl} 
\usepackage[table]{xcolor} 
\usepackage{booktabs} 
\usepackage[capitalise]{cleveref}
\usepackage{url}
\usepackage{subcaption}

\begin{document}

\title{A Critical Discourse Analysis of Gender Representation in Software Engineering Education Videos on YouTube}
\titlerunning{Gender Representation in SE Education Videos on YouTube}

\author{Isabella Graßl \inst{1}\orcidID{0000-0001-5522-7737} \and
Alexander Serebrenik\inst{2}\orcidID{0000-0002-1418-0095} \and
Giuseppe Destefanis\inst{3}\orcidID{0000-0003-3982-6355}}
\authorrunning{Graßl et al.}
%
\institute{Technical University of Darmstadt, Germany \and
Eindhoven University of Technology, Netherlands \and
University College London, UK
}

\maketitle

\begin{abstract}
Educational resources may frame students' perceptions of who belongs in software engineering, which is relevant given the field's ongoing gender gap. However, we know little about the \emph{hidden curriculum} regarding gender in online learning spaces.
This study presents a critical discourse analysis of 200 manually analysed English and German software engineering tutorials on YouTube, examining gender representation through contextual domains and linguistic identity markers.
Our results show that male characters and masculine linguistic defaults dominate the tutorials. We identified an \emph{agency gap}, in which technical and decision-making roles are almost exclusively assigned to male actors, while female actors are either absent or tend to passive, low-agency roles.
The findings indicate that software engineering education on YouTube may reproduce gendered norms, in which linguistic and representational gatekeeping may serve as a symbolic barrier to software engineering.


\end{abstract}

\section{Introduction}
The way we represent people in educational materials matters as it frames how learners are introduced to the practices and identities of a discipline~\cite{master2016,spieler2020,shardakova2004}. 
In software engineering, educational resources do more than just illustrate technical concepts as they guide students towards specific ways of thinking about roles, problems, and the people who solve them~\cite{kovaleva2022c,svedin2016,kovaleva2022d}. 

These \emph{hidden curricula}~\cite{snyder1971,jackson1968} in human examples can communicate unspoken rules about belonging and  identity~\cite{logacheva2024,kong2022,santhosh2024}. This is important because they can either contribute to the lack of a sense of belonging among female students in software engineering~\cite{boman2024} or help tackle stereotypes around technical identity~\cite{chattopadhyay2021a}.

While research has explored gender\footnote{By gender, we refer to the socio-culturally constructed roles and identities.} stereotypes in sense of belonging in software engineering~\cite{boman2024}, in computing textbooks for mostly children~\cite{dewit2024,grassl2025b,dele-ajayi2020a}, and what kinds of work and materials are valued~\cite{svedin2016,santhosh2024}, a gap remains in understanding how these norms are embedded in online spaces like YouTube. YouTube has an impact on how software engineering can be perceived as a profession~\cite{chattopadhyay2021a}, and has become a valuable resource for students in technical fields~\cite{engelmann2024}. This is relevant regarding cross-linguistic interactions with grammatical structures such as the German generic masculine\footnote{The German generic masculine refers to the traditional grammatical rule where masculine plural forms or titles are used to refer to groups of mixed or unknown gender (e.g., using \emph{der Entwickler} for any developer).}~\cite{dewit2024,rothermund2024,engelmann2024}. 
%

To address this, we conducted a novel exploratory study using \emph{Critical Discourse Analysis} (\emph{CDA})~\cite{fairclough2013} to analyse 200 manually coded YouTube tutorials in German and English. We address the following research question:\\

\noindent
\textbf{RQ:} How is gender represented within software engineering tutorials on YouTube across different linguistic contexts?\\

Our results show that male actors are far more represented than female actors within software engineering education on YouTube. We found an \emph{agency gap}, in which high-power technical roles are consistently coded as male, particularly in German-language content, whereas female actors tend to be assigned to non-technical, low-agency roles. While the English sample showed more surface-level linguistic neutrality, both languages struggle with a hidden curriculum that associates technical authority with masculinity.

These findings suggest that the way software engineering is taught and represented online may reinforce already negative associations with the male-dominated field by signalling to non-male learners that they do not belong. 
\section{Background and Related Work}
Software engineering reflects the values, biases, and perspectives of its human creators, often privileging a male-centric perspective~\cite{boman2024}, making it an important subject for research through socio-cultural and gender lenses~\cite{hermans2024,stattkus2025,bhargava2002}. If the discipline is perceived as \emph{dude culture}~\cite{miller2021}, this can manifest in software engineering education courses and material~\cite{kovaleva2022c,kovaleva2022d}, creating barriers to entry and affects who feels they belong in the software world~\cite{dele-ajayi2020,dewit2024,medel2017b}.

\subsection{The Hidden Curriculum and Software Engineering}
A key concept for understanding these barriers is the \emph{hidden curriculum}. This refers to the implicit messages, values, and norms that are taught through the choice of examples, language, and classroom culture, often without being part of the formal syllabus~\cite{snyder1971,jackson1968}. While not necessarily intentional, the hidden curriculum communicates who is considered a \emph{natural} fit for the field~\cite{nakai2023,martin1976}.

Even in subjects with high diverse gender representation such as social sciences, the inclusion of diverse identities in textbooks is limited~\cite{smestad2018,tsui2009,dele-ajayi2020a}. 

In software engineering, this is particularly evident in how we present subfields of such as requirements analysis, software modeling, and project management. The choice of domains and its human actors in its examples reflects assumptions about which problems are worth solving and who the participants are~\cite{lin2022,treude2023,vasilescu2014}. When educational materials consistently use stereotypically male names, Western contexts, or heteronormative roles, they subtly reinforce the image of a \emph{typical} professional~\cite{santhosh2024,nakai2023}. This can discourage students who do not see themselves in these materials, ultimately shaping their computing identity and career aspirations~\cite{santhosh2024,svedin2016}.
This also matters beyond education, since gender diversity has also been linked to socio-technical communication patterns in software teams~\cite{catolino2019}.

\subsection{Representation from Textbooks to YouTube}
The study of bias in educational materials has traditionally focused on textbooks. Research across various cultures and subjects, including mathematics in Germany and chemistry in the US, consistently shows that male characters appear more frequently and are more often portrayed in active, professional roles, while female characters are relegated to passive or domestic positions~\cite{chi2024,murray2022}. In computer science specifically, biases often appear in naming conventions (like \emph{Alice and Bob}), the trope of the \emph{lone male programmer}, or male-associated topics~\cite{medel2017b,dewit2024,grassl2025b}. While some research has analysed the technical accuracy or computing careers of vlogs and TikTok~\cite{chattopadhyay2021a,kadriu2020,martinez2026}, there is a gap in understanding the \emph{social} and \emph{cultural} messages these videos send and how they reproduce~\cite{martinez2026}, or challenge~\cite{chattopadhyay2021a}, gender stereotypes. 

Some research demonstrates the importance of examining these materials critically: Engelmann et al.~\cite{engelmann2024} found that the most watched introductory AI courses on YouTube fail to meaningfully engage with ethical or societal challenges, while other research~\cite{abdelhamid2021} highlighted the role of figurative language in shaping how learners perceive computing tutorials.

\subsection{Critical Discourse Analysis}
\emph{Critical Discourse Analysis (CDA)} can be used to examine how power, dominance, and inequality are produced and resisted through language and visual communication~\cite{fairclough2013}. Originating from critical theory, \emph{CDA} looks at \emph{how} actors are spoken to and about by analysing the \emph{discourse}, the combination of words, tone, and imagery, to see how it constructs reality.

\emph{CDA} can be particularly useful for software engineering research because software engineering is built on communication, where we look at how the language used in a YouTube tutorial position software engineering. This lens allows us to move beyond the surface level to understand how educational content creates an environment that either welcomes or excludes certain groups, providing a deeper look at the social structures embedded in technical education.

To our knowledge, there has been no empirical study on gender representation in software engineering education materials on YouTube and particularly not within a socio-cultural framework like \emph{CDA}.

\section{Methods}
We use a qualitative approach to examine how human identities are represented in software engineering tutorials on YouTube. Our research followed a multi-stage process. We built a large dataset to allow for future replication, and then we used stratified sampling to create a representative subsample, which provided us with a manageable group of videos for detailed manual coding.

\subsection{Data Collection and Dataset}
\subsubsection{Collection Pipeline}
The dataset was built using a Python script that integrates the \texttt{youtube-transcript-api} library and the YouTube Data API. The pipeline proceeds through the following steps:

\emph{(1)} A predefined list of search terms was compiled in multiple languages, targeting concepts from software engineering and modelling. These included terms such as ``UML class diagram tutorial'', ``sprint planning example'', and ``build a simple web app'', along with their translated equivalents.\footnote{We provide all search terms: \url{https://figshare.com/s/d6cc8319e2ad2f639546}}
    
\emph{(2)} For each search term across languages, the YouTube Data API was used to retrieve up to fifty videos returned by the YouTube search endpoint. 

\emph{(3)} For each retrieved video, metadata was extracted, e.g., the title, channel, publication date, duration, and tags.

\emph{(4)} The transcript was retrieved using the \texttt{youtube-transcript-api}, which supports access to both manually uploaded and automatically generated transcripts. Only videos with a valid (i.e., manually checked) transcript were retained, and the transcript type was recorded. 

\emph{(5)} Data was stored using the \texttt{pandas} library and saved in CSV format. Basic cleaning steps included standardising date formats, and ensuring the completeness of metadata fields.

Our pipeline is modular and supports future extension by adding search terms in additional languages or modifying filtering criteria. This structure allows for scalable multilingual collection with minimal manual intervention.

\subsubsection{Inclusion Criteria}
Videos were included in the dataset based on the following criteria. First, the video had to be publicly accessible on YouTube at the time of collection (Sept. 2025) and returned among the top fifty results for a relevant search term. The search terms were chosen from discussions among the authors to cover foundational topics in software engineering education, including but not limited to software modelling (e.g., class diagrams, use case diagrams), development processes (e.g., agile, scrum, sprint planning), programming tutorials (e.g., JavaScript projects, Python for beginners), and team-based interactions (e.g., client meetings, user story mapping). 
For each target language, search terms were translated from English to German and Italian by the authors, who are fluent or native speakers. The translations aimed to reflect commonly used expressions in educational content specific to each language community.

Second, the video content had to be pedagogical in nature. This was verified by manually inspecting titles, descriptions, and a sample of the transcript. Promotional content, tool advertisements without educational value, and non-technical videos were filtered out.

Finally, videos were required to have a minimum duration of one minute. This threshold was selected to exclude clips that are too short to have \emph{meaningful} instructional content, such as trailers, or promotional ads. While content on platforms like TikTok or YouTube Shorts can be used as informal resource~\cite{martinez2026}, a minimum of one minute was considered necessary to allow the presence of a complete and self-contained explanation or example~\cite{engelmann2024}.

\subsection{Dataset and Stratified Sample}

\subsubsection{Dataset} 
The dataset comprises 1,225 YouTube videos, including videos in English ($n=702$), Italian ($n=345$), and German ($n=178$). The difference in the number of videos per language is primarily due to the size of the respective YouTube community since English is the most common language on the platform, resulting in a much larger volume of educational content. 

The dataset is stored in a structured CSV file containing core metadata for each tutorial. This includes identifying information such as the \emph{video\_id}, \emph{title}, and \emph{channel}, alongside the \emph{search\_term} used for retrieval. To support our analysis of content for future work, we recorded the \emph{view\_count}, \emph{like\_count}, \emph{duration}, and \emph{tags}. The dataset preserves the full \emph{transcript} and identifies the specific \emph{language} and \emph{publication date} for each entry.

To provide a clearer view of the dataset’s structure, we grouped video durations into categories based on standard content segmentation practices observed in online educational media. The intervals reflect distinctions between short-form, medium-form, and long-form instructional content. The categories are: under 5 minutes, 5–10 minutes, 10–20 minutes, 20–30 minutes, 30–60 minutes, and over 60 minutes.

\subsubsection{Stratified Sample} 
To ensure the reliability of our manual content analysis regarding gender bias, we determined the sample size through statistical power analysis.  
For this study, we focused our stratified sample on English and German and excluded Italian videos. This was due to a practical limitation: the researcher responsible for the Italian analysis was unavailable during the manual coding phase. However, we have kept the 345 Italian videos in our main dataset and provided them in our replication package so that other researchers can include them in future studies.

We applied standard sample size calculations with finite population correction, adopting a conservative expected proportion of biased videos to maximise variance and ensure adequate power. To achieve 95\% confidence with a ±5\% margin of error, we identified specific minimum sample requirements for each language group. For the English population, a minimum sample of 239 videos is required. The German dataset requires 122 videos, respectively, to maintain statistical significance. These calculated thresholds ensure that our qualitative analysis remains representative of the broader corpus.

For practical feasibility and robust cross-language comparisons, we implemented equal allocation, selecting 100 videos per language (total $n=200$). This stratified random sample was proportionally distributed across video durations, ensuring both short- and long-form content were represented.  

This allocation ensures (1) an equal statistical power across languages for cross-group comparisons, (2) adequate coverage of different video lengths, accounting for potential exposure to gendered content, and (3) a manageable workload for detailed manual coding.

Our final sample includes videos from 2010 to 2025, where the majority are most recent videos (2025--2020: 66\%, 2019--2015, 26.5\%). The average number of view is 470,535,95 for English videos and 51,237,54 for German ones. Since most of the videos are recent that implies that our analysis primarily reflects current platform dynamics and discourse, rather than long-term historical trends.

\subsection{Gender Evaluation Framework}
\label{sec:framework}
Our framework (\Cref{tab:framework}) is grounded in \emph{Critical Discourse Analysis (CDA)} to examine how technical tutorials construct gendered identities, representations, and discursive agency.

\begin{table}[t]
\centering
\tiny
\caption{Analytical framework for gender representation.}
\label{tab:framework}
\begin{tabular}{p{3.2cm}p{9.2cm}}
\toprule
\textbf{(Sub-)Category} & \textbf{Description} \\
\midrule
\multicolumn{2}{l}{\textbf{Contextual Domain}} \\

Business / Commercial& Scenarios involving commerce, banking, shopping, or corporate management. \\
Technical Only & Abstract programming concepts without human or social context. \\
Social / Entertainment & Community-focused contexts, gaming, or social media. \\
Public / Education& Healthcare, schools, or public-good scenarios (e.g., environmental data). \\
\midrule
\multicolumn{2}{l}{\textbf{Identity Markers}} \\

Named Characters & Specific individuals (e.g., \emph{Bob}, \emph{Alice}) implying gender. \\
Role References &  Pronouns linked to professional titles (e.g., \emph{the developer} $\rightarrow$ \emph{he}).\\
Generic / Default & Pronominal choices for unspecified identities. \\
Role Agency & Distribution of activation (High Agency: \emph{Manager}) vs passivation (Low Agency: \emph{User}). \\

\midrule
\multicolumn{2}{l}{\textbf{Bias Indicators}} \\

Male-Centric Technicality & Erasure of non-male actors in all expert or developmental roles. \\
Generic / Default & Use of masculine forms as the standard. \\
Passive Roles & Assignment of female actors to support, consumer, or domestic roles. \\
Stereotypical roles& Assignment of gender stereotypical roles.\\

\midrule
\multicolumn{2}{l}{\textbf{Representation Index}} \\
Balanced & Use of balanced role distribution. \\
Minor bias  & Occasional gender skew or informal address.\\
Clear bias  & Systemic reproduction of stereotypes. \\
Strong bias  & Systemic exclusion or discriminatory language. \\
\bottomrule
\end{tabular}
\end{table}

\emph{Domain. }
We categorise the settings of examples within the tutorials because they reflect the \emph{field} of discourse, i.e., the social or professional field in which technical examples are situated. 
Based on Fairclough’s argumentation~\cite{fairclough2005}, the domain (e.g., Business vs. Education) controls which language is considered standard and who is empowered to speak. 

The \emph{Technical} domain covers pure programming concepts and architecture, while \emph{Business/Commercial} contexts focus on market-oriented applications. \emph{Social/Entertainment} domains focus on community and leisure, and \emph{Education/Public} contexts include university settings, healthcare and public services.


\emph{Identity Markers. }
This dimension examines how human identities are constructed through linguistic choices~\cite{medel2017b}. 
We treat these cues as markers of gendered representation rather than as evidence of actual gender identity. This distinction is important because prior research~\cite{serebrenik2024} has shown that inferring gender from archival traces such as names, profiles, or visual cues is inherently uncertain and can reproduce binary assumptions about gender.
Following Van Leeuwen’s theory of social actor representation~\cite{leeuwen2008}, we distinguish between \emph{Activation} (high-agency) and \emph{Passivation} (low-agency). 

High-agency roles represent actors with technical power, such as an \emph{Admin} or \emph{Architect}, while low-agency roles represent those who receive technology, such as \emph{Patients} or \emph{Customers}. We also analyse \emph{nomination strategies}, i.e., how people are named~\cite{reisigl2017}. 
For example, choosing a gender-ambiguous name like \emph{Alex} suggests a different discursive intent than consistently using binary names like \emph{Bob} or \emph{Alice}. 
Furthermore, we track how pronouns indicate gender in roles such as \emph{the developer}; in German, this often defaults to the generic masculine, whereas English may use the neutral \emph{they}. A common marker of bias here is when the \emph{Lead Engineer} is consistently referred to as \emph{he}, while the \emph{User} is referred to as \emph{she}, regardless of the code's function.

\emph{Bias Indicators. }
We investigate the implicit gender norms and socio-cultural assumptions embedded in the examples~\cite{paiz2015}. 
These indicators are patterns and markers of internalised ideologies, stereotypes, and socio-cultural norms that reinforce a \emph{male-as-default} ideology. 

This includes \emph{Male-Centric Technicality}, where every technical expert shown is male, and the \emph{Generic Masculine Default}, where masculine pronouns are used for any unknown gender. We examine \emph{Passive Role Assignment}, in which female actors appear only in supporting roles. 
In addition, we look for \emph{Stereotypical role assignments}. A common example from the  literature is the \emph{Doctor/Nurse} trope, which, in a software engineering context, translates to the \emph{Systems Architect} being male and the \emph{Designer} or \emph{Customer Support} being female.

\emph{Gender Representation Index. } 
The final dimension of our framework assigns a score reflecting the overall discursive content of the tutorial, following prior work~\cite{bhargava2002}. 
This index ranges from \emph{Balanced} content to \emph{Strong Bias}. To illustrate the difference: a video about a database system in which one example names an \emph{Admin} \emph{Bob} constitutes \emph{Minor Bias} (Score 1). If there are multiple occurrences of \emph{Bob} or stereotypical role assignments, this constitutes a structural exclusion that constructs a male-only professional reality through repetition, resulting in a \emph{Clear Bias} (Score 2). In contrast, \emph{Strong Bias} (Score 3) would involve explicitly problematic content, such as a creator making a joke about women’s technical ability. This distinction allows us to separate unintentional structural habits from active gatekeeping.

\subsection{Data Analysis}

\paragraph{Qualitative Content Analysis}
We used the subset from the stratified sampling, including 100 German videos and 100 English videos. In total, this amounts to over \emph{94 hours} of content to read, watch for visual cues, and analyse.

We followed Mayring’s approach to qualitative content analysis~\cite{mayring2014}. We used deductive coding to apply the categories from our analytical gender evaluation framework to the videos. 
In a first round, two researchers independently coded ten English videos according to the framework. This resulted in an inter-rater agreement of Cohen’s $\kappa = .87$, indicating almost perfect agreement. Since this level of agreement suggests a high degree of coding reliability, the first author continued with the remaining analysis independently. In four cases, the author engaged in in-depth discussions with a research assistant to clarify borderline cases and ensure consistent interpretation of the coding scheme. 
In the process, we took notes on recurring patterns, such as how characters were named or the specific roles they played within certain domains. We also wrote a brief explanation for each bias score to capture the rationale behind it. 

\emph{Domain. }
We identified the domains used by creators in their examples, looking at both the (spoken) text and visual elements. Each video could belong to one or more domains. While we started with the pre-defined domains in our framework, we allowed ourselves to add new ones if they appeared in the data. We also recorded combinations when a video used multiple examples from different areas. For each domain or combination, we report the relative frequency to show how these settings are distributed across our sample.

\emph{Identity Markers. }
To understand gender visibility, we used a weighted counting method. Our raw data consisted of gender combinations, e.g., \texttt{2m1f0n} denotes two males, one female, and zero neutral entities. We also recorded the frequency with which each combination appeared.
Instead of just counting rows in our spreadsheet, we calculated the total number of individual entities. We did this by multiplying the number of people in each combination by the frequency with which that combination occurred. If the \texttt{2m1f0n} group appeared five times, we recorded 10 male and 5 female entities. This ensures that a video with many references has more weight in our final percentages than a video with only one.

To understand social grouping styles, we classified these occurrences into four types: \emph{Pure Male / Female / Neutral}, for groups in which only one gender is present, and \emph{Mixed}, for groups in which two or more genders co-occur.

We also looked beyond linguistic differences, since German often relies on the generic masculine, whereas English uses the neutral \emph{they}; comparing them directly can be difficult. To address this, we performed a separate analysis in which we isolated the male ($m$) and female ($f$) entities. By removing the neutral ($n$) data points, we could calculate a Male-to-Female ratio and compare representation across both languages on more equal terms.

\emph{Bias Indicators. }
For each indicator subcategory, we made a binary decision: did the video contain this specific bias, e.g., are all technical roles in the video assigned to male actors? We counted the \emph{yes} and \emph{no} decisions for each subcategory and reported these as relative values (percentages) for each language.

\emph{Gender Representation Index}
Our index ranges from -1 (inclusive content) to 3 (strongly biased), following a checklist of gender bias in teaching materials~\cite {bhargava2002}. We calculated the average score for each language and identified the highest and lowest scores in the dataset. We also applied content analysis to the written explanations for these scores. This enabled us to identify shared reasons for bias and to select representative examples to illustrate our findings.

\subsection{Researcher Positionality. }
Our research team consists of people of various genders based in different European countries, and with several years of experience in software engineering education as well as qualitative analysis. We identify as members of the queer community, which provides us with certain lens through which we might view representation and inclusion in software engineering. 

We acknowledge that our backgrounds may influence how we conduct research and interpret data. However, in \emph{CDA}, the researcher is not a neutral observer but an active participant in the interpretation process~\cite{reisigl2017}. 
To ensure the reliability of our findings, we used a review process in which team members discussed the coding and interpretations. 

\textbf{Ethics.} Our study relied on publicly available YouTube content with an official API. We have discussed the sensitive topics and agreed that we focus on the discursive patterns of the content rather than blaming individual creators. We accessed and analysed the video data in accordance with the platforms' terms of service and standard ethical guidelines for research.

\textbf{Data Availability.} To support verification and replication, we share the whole dataset as well as the stratified sampling, including the codebook from the qualitative content analysis: \url{https://figshare.com/s/d6cc8319e2ad2f639546}.

\subsection{Threats to Validity}
\emph{Construct validity.} Our framework and coding system might not fully capture how people are represented in the video tutorials. To address this, we used data triangulation by combining pre-defined categories from prior literature with manual descriptions of concerns or problematic issues in the video content.  
Another potential issue is the number of videos, since we simply cannot manually cover all the content on YouTube regarding tutorials. However, we used stratified sampling to get a representative subset, and manually viewed the included videos. Finally, a related construct threat concerns gender coding itself (cf.\ \cite{serebrenik2024}).

\emph{Internal validity}: Since we coded the videos manually, our personal background could affect the interpretation. We tried to minimise this by creating a detailed, shared codebook that we also share for replication. When we were unsure about how to categorise a specific example, we discussed it as a group. 

\emph{External validity}: Our study is limited to YouTube videos in German and English. They do cover a large global community, but do not represent all cultural or linguistic contexts. Therefore, our findings may not apply to software engineering education in other languages or on other platforms.

\section{Results}
We present qualitative content analysis of 200 YouTube videos in German and English according to the  gender evaluation framework in \Cref{sec:framework}. 

\subsection{Domain}\label{sec:domain}
\begin{figure}[t]
    \centering
    \includegraphics[width=0.5\columnwidth]{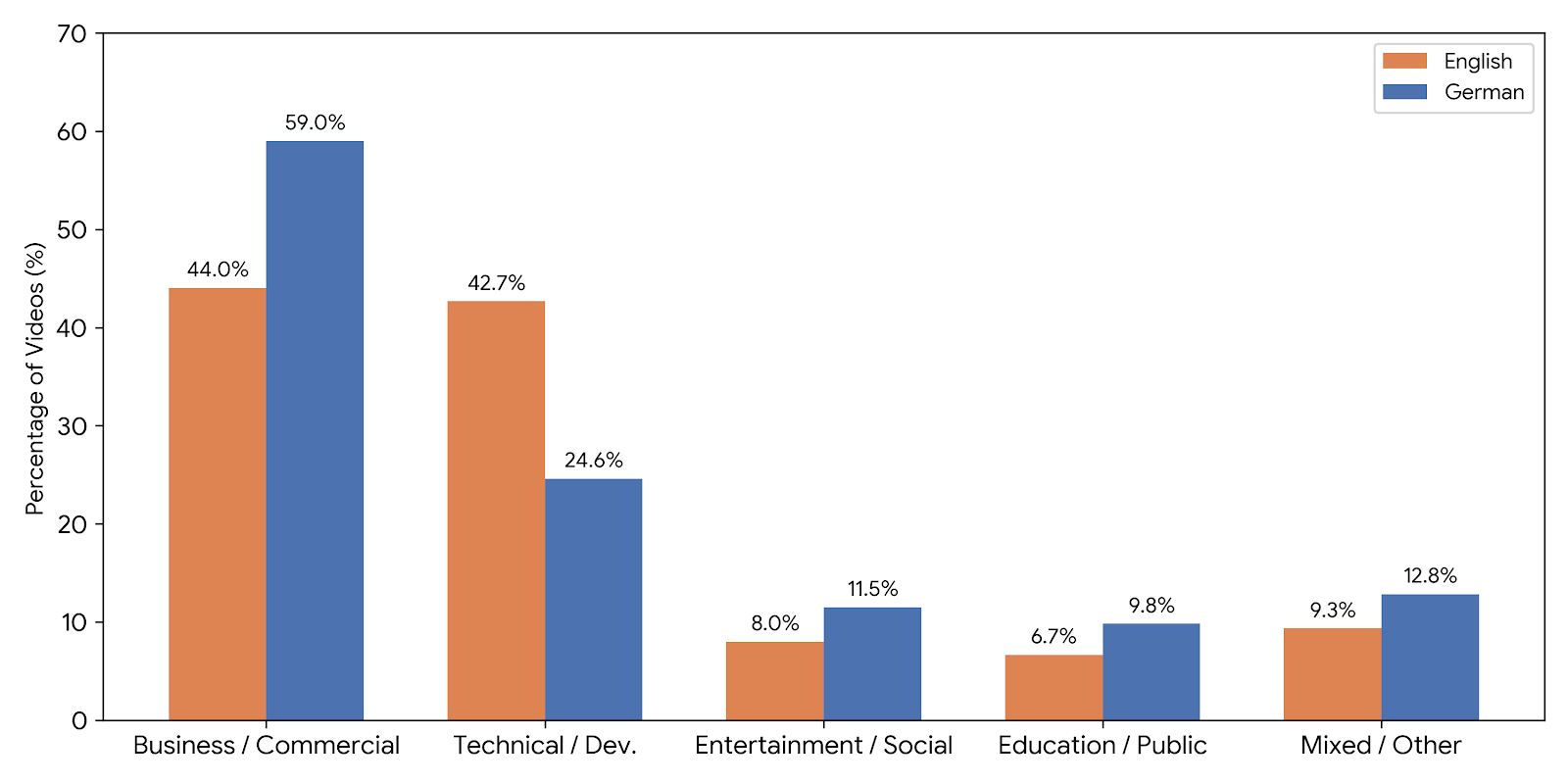}
    \caption{Distribution of context domains across videos.}
    \label{fig:domain}
\end{figure}
The \emph{Domain} category identifies the situational context in which examples are placed. 
\Cref{fig:domain} illustrates that the primary setting for software engineering tutorials differs between the two languages.

In the German sample, the Business/Commercial domain is the dominant setting, accounting for 59.02\% of all manually analysed videos, which is followed by Technical/Software Development at 24.59\%. 

The English sample presents a more balanced two-pillar approach. While the Business/Commercial domain is still the largest single category at 44.00\%, the Technical/Software Development domain follows very closely at 42.67\%. 

Both languages introduce videos with broader multi-domain combinations, such \emph{Business \& Entertainment \& Education} (2.67\%). 

This implies that German software education is frequently framed through the lens of commerce and industry, while English education is more frequently framed through both lenses of the craft of business and software itself. Both languages maintain a similar, low level of focus on Education/Public and Entertainment domains, suggesting these are underutilised as pedagogical contexts.

\subsection{Identity Markers}\label{sec:identity}
\begin{table}[t]
\centering
\tiny
\caption{Identity markers from the gender evaluation framework for our sample.}
\label{tab:comprehensive}
\begin{tabular}{l l | cccc | ccc | ccc}
\toprule
\textbf{Category} & \textbf{Language} & \multicolumn{4}{c}{\textbf{Total Videos}} &
\multicolumn{3}{c}{\textbf{Count entities}}&
\multicolumn{3}{c}{\textbf{Percentage}}
\\
& & \textbf{Pure M} & \textbf{Pure F} & \textbf{Pure N} & \textbf{Mixed} & \textbf{M} & \textbf{F} & \textbf{N} & \textbf{\% M} & \textbf{\% F} & \textbf{\% N} \\ \midrule
Named Characters & German & 19 (4) & -- & -- & 4 (4) & 37 & 8 & 1 & 80.4\% & 17.4\% & 2.2\% \\
 & English & 8 (2) & 2 (1) & -- & 4 (4) & 16 & 9 & 1 & 61.5\% & 34.6\% & 3.8\% \\ \midrule
Role References & German & 35 (9) & -- & 1 (1) & 19 (18) & 188 & 11 & 40 & 78.7\% & 4.6\% & 16.7\% \\
 & English & -- & -- & 53 (9) & 6 (6) & 10 & 7 & 236 & 4.0\% & 2.8\% & 93.3\% \\ \midrule
Generic/Default & German & 29 (7) & -- & -- & 22 (18) & 184 & 8 & 52 & 75.4\% & 3.3\% & 21.3\% \\
 & English & 1 (1) & -- & 42 (9) & 7 (7) & 15 & 7 & 209 & 6.5\% & 3.0\% & 90.5\% \\ \midrule
High Agency & German & 24 (5) & -- & -- & 17 (16) & 125 & 12 & 29 & 75.3\% & 7.2\% & 17.5\% \\
 & English & -- & -- & 34 (6) & 9 (8) & 15 & 8 & 140 & 9.2\% & 4.9\% & 85.9\% \\ \hline
Low Agency & German & 29 (6) & -- & -- & 14 (7) & 92 & 5 & 24 & 76.0\% & 4.1\% & 19.8\% \\
 & English & 1 (1) & -- & 37 (6) & 10 (8) & 15 & 8 & 107 & 11.5\% & 6.2\% & 82.3\% \\ 
\bottomrule
\end{tabular}
\end{table}

\begin{table}[t]
\centering
\tiny
\caption{Representation Excluding Neutral/Generic Language.}
\label{tab:binary}
\begin{tabular}{l | cc | cc}
\toprule
\textbf{Category} & \multicolumn{2}{c|}{\textbf{German M:F Ratio}} & \multicolumn{2}{c}{\textbf{English M:F Ratio}} \\ 
 & Ratio & \% Female & Ratio & \% Female \\ \midrule
Named Characters & 4.6 : 1 & 17.8\% & 1.8 : 1 & 36.0\% \\
Role References & 17.1 : 1 & 5.5\% & 1.4 : 1 & 41.2\% \\
High Agency & 10.4 : 1 & 8.8\% & 1.9 : 1 & 34.8\% \\
Low Agency & 18.4 : 1 & 5.2\% & 1.9 : 1 & 34.8\% \\
 \bottomrule
\end{tabular}
\end{table}

\begin{figure}[t]
    \centering
    \subfloat[
        Gendered naming patterns in its example dataset ({\tiny left; fgOiWEGNJ-o}), or a specific student ({\tiny right; pPaVvPJ37YM}).
        \label{fig:negative}
    ]{
        \includegraphics[width=0.42\columnwidth]{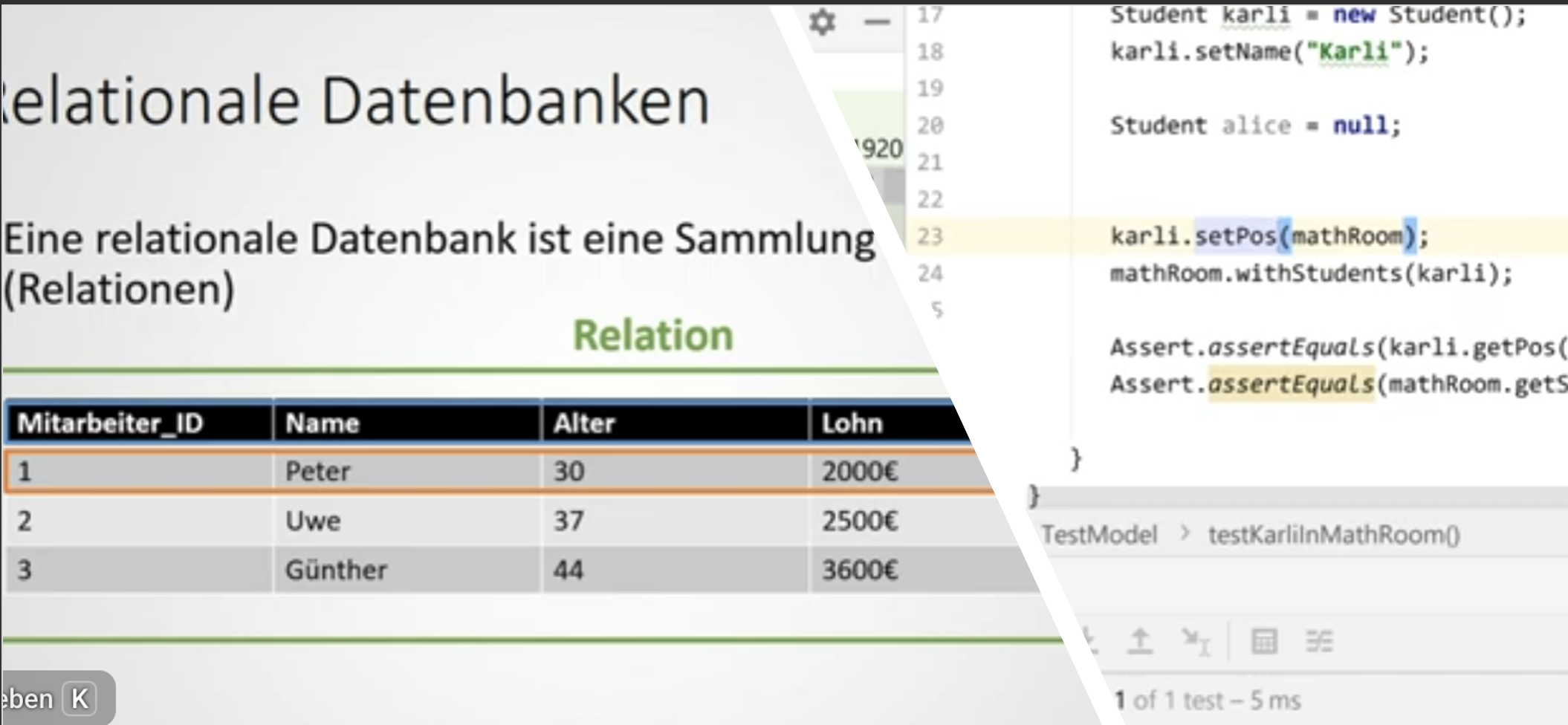}
    }
    \qquad
    \subfloat[
        Naming and agency {\tiny left; k-oSAeXWZYY}) or neutral figures with gender-sensitive linguists ({\tiny right; 7qiuGST7EVQ}).
        \label{fig:mixed}
    ]{
        \includegraphics[width=0.42\columnwidth]{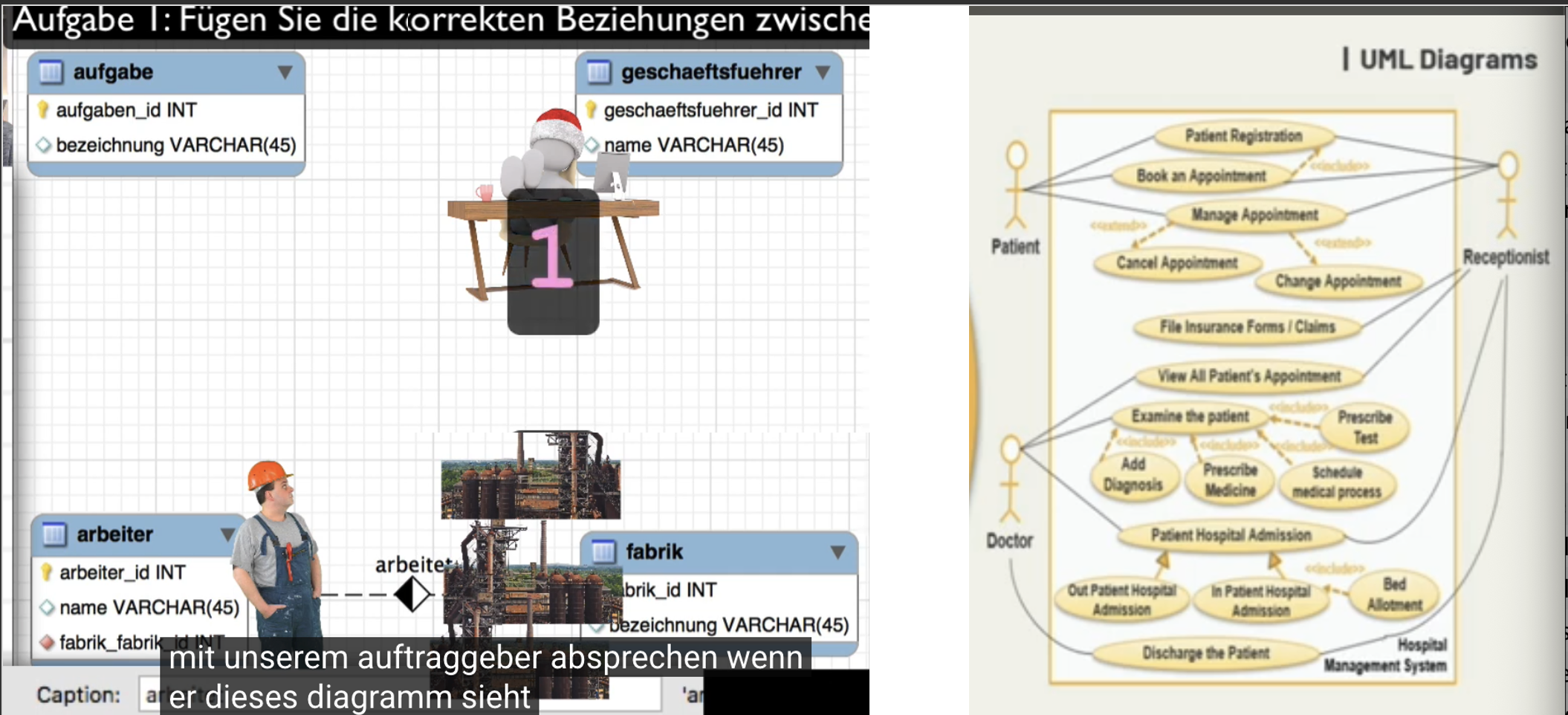}
    }
\end{figure}
Identity markers are specific linguistic and visual cues, such as names, pronouns, and titles, used to assign gender to human actors in a tutorial. 
\Cref{tab:comprehensive} shows the combination styles and absolute entity counts, while \Cref{tab:binary} isolates the data by removing neutral markers to highlight the underlying social representation.

\subsubsection{Named Characters} \label{sec:names}
Creators on YouTube personify their examples with specific names more often in German than in English, though both languages exhibit a clear male-name dominance. 
Since names typically imply a specific gender, neutral language is almost non-existent here ($<4\%$), making this category a baseline for social representation independent of grammar.

In the German sample, we find 19 \emph{Pure Male} videos, meaning only male names occur. The most common combination is \texttt{1m0f0n}, which appears in 14 videos, whereas larger groups, such as \texttt{4m0f0n}, are rare. Mixed-gender groupings occur in only four videos, and \emph{Pure Female} representation is limited to a single instance. This results in males accounting for over 80\% of all named entities in German tutorials.

One German Video, \texttt{fgOiWEGNJ-o}, presents three distinct technical scenarios. In all three scenarios, every character is male, and the language remains strictly masculine, leaving no space for female representation (\Cref{fig:negative}). 
In another example, \texttt{pPaVvPJ37YM}, the creator uses a single example for coding, where it is a named student, \emph{Karli}, in a maths course (\Cref{fig:negative}). Here, the contextual domain represents maths as commonly perceived: male and complex, with a male named. In video \texttt{k-oSAeXWZYY}, a construction worker is referred to as \emph{Heini} (a colloquially masculine German term, an abbreviation of the male name Heinrich), and all accompanying visual aids use exclusively male images alongside generic masculine language. This is especially interesting as it also shows the CEO (\emph{Geschäftsführer} male, but with a more \emph{neutral} look (\Cref{fig:mixed}). 
We see such combinations in several videos, e.g. popular male actors such as \emph{Thomas Müller}, who is a popular German football player or fictitious \emph{Tony Stark} from Iron Man as a genial mad technocrat.

The English sample shows a slightly more diverse distribution. Although \texttt{1m0f0n} is still the most frequent single pattern (appearing in seven videos), we see \emph{Pure Female} combinations in two videos (each video having \texttt{0m1f0n}).
We also identify 34.6\% of the named population is female. This is the closest the data comes to gender parity in any category.
Compared with the German video topic, we observed one video showcasing an implementation of a shopping website with both a women's and a men's section, with a focus on the latter, rather than going back to female appearance and shopping stereotypes (\texttt{8bkGKwb29L4}). 

When we compare the two languages, for every woman named in a German tutorial, there are 4.6 men, compared to 1.8 men in English (\Cref{tab:binary}). This suggests that English creators are more likely to include women as specific, named individuals in their examples of software engineering education.

\subsubsection{Role References}
Role references (e.g., \emph{the developer}) reveal the most significant split in the dataset. While creators in both languages frequently refer to roles such as \emph{the developer} or \emph{the programmer}, their choice of pronouns depends almost entirely on the language spoken. This category contains nearly 500 total entity references (\Cref{tab:comprehensive}).

The German data are male-dominated, with 35 videos that use exclusively male pronouns for referencing any roles. The \texttt{2m0f0n} combination is the most frequent pattern (13 times). We also observe highly skewed \emph{Mixed} outliers, such as the \texttt{9m0f2n} combination, which uses numerous male references within a single context. Women are mentioned as roles in only 4.6\% of cases, typically within mixed groups.
In contrast, English tutorials are almost entirely neutral, with 53 \emph{Pure Neutral} videos using the singular \emph{they} as an indefinite or gender-neutral pronoun~\cite{saguy2022little}. When English speakers do use a gendered pronoun, they are relatively balanced; female references account for 41.2\% of non-neutral instances. 
The odds of a role being referred to as male are 17 times higher in German.

\subsubsection{Generic and Default Language}
This category describes hypothetical users or generic subjects. It highlights representational erasure, where female visibility drops compared to when specific people are named.

The German sample relies on \emph{Pure Male} patterns (29 videos) and \emph{Mixed} groupings (22 videos). In these mixed groups, the male count often vastly outweighs the others (e.g., \texttt{11m0f1n}). There are no instances of purely female or purely neutral groupings in this category for German, resulting in a 75.4\% male entity share.

English tutorials are almost entirely neutral here, with 42 videos using \emph{Pure Neutral} language. The most common pattern is \texttt{0m0f1n} (seven videos), although we also observe high-frequency patterns such as \texttt{0m0f4n} (eight videos). Male and female linguistic clues appear only rarely and usually in small, mixed amounts with instances where people intentionally \emph{his} or \emph{her} (e.g. \texttt{SS3k1X6r7s0}). However, in one English tutorial, the creator repeatedly addresses their audience as \emph{bros} and encourages viewers to subscribe to become a \emph{fellow bro}. This creates a clear discursive boundary excluding anyone who does not identify as a \emph{bro}~\cite{wang2024}.

The comparison shows that the \emph{erasure} of women in linguistics is far more prominent in German. When moving from specific names to generic defaults, female visibility in German decreases by 13\%, whereas in English it decreases by about 4\%, which indicates that the generic masculine in German might function as a barrier to female visibility.

\subsubsection{Role Agency}
The agency analysis examines whether characters are represented as active leaders (High Agency) or passive recipients (Low Agency). Overall, men are more often portrayed as active leaders of a task (\Cref{tab:comprehensive}). 

In the German sample, men dominate both High and Low agency roles. In High Agency, we observe combinations with groups of men, such as \texttt{9m0f1n} and \texttt{8m1f0n}, with women attaining 7.2\% visibility. In Low Agency, women drop to 4.1\%, and the \texttt{2m0f0n} pattern appears most frequently (10 times), indicating that men are the default for every task.

In the English sample, agency does not appear to be gendered that strongly. 
We found no \emph{Pure Male} videos in High Agency, and the most common pattern across both categories was neutral (e.g., \texttt{0m0f2n} for Low Agency in 11 videos). 
There are several examples that explicitly use gendered forms for high-agency roles (e.g., \texttt{gW1g-bq8c8o}). One English video explicitly uses \emph{him} or \emph{her} for high-agency roles, such as doctors in a hospital management system (\Cref{fig:mixed}).

In German, the male-to-female ratio for Low Agency roles (18.4:1) is nearly double that of High Agency roles (10.4:1, \Cref{tab:binary}), indicating that women in German software education are less likely to even be used as placeholders for passive technical examples.

\subsection{Bias Indicators} \label{sec:red-flags}
\begin{figure}[t] 
\subfloat[ Comparison of bias indicators. \label{fig:red-flags} ]{ \includegraphics[width=0.44\columnwidth]{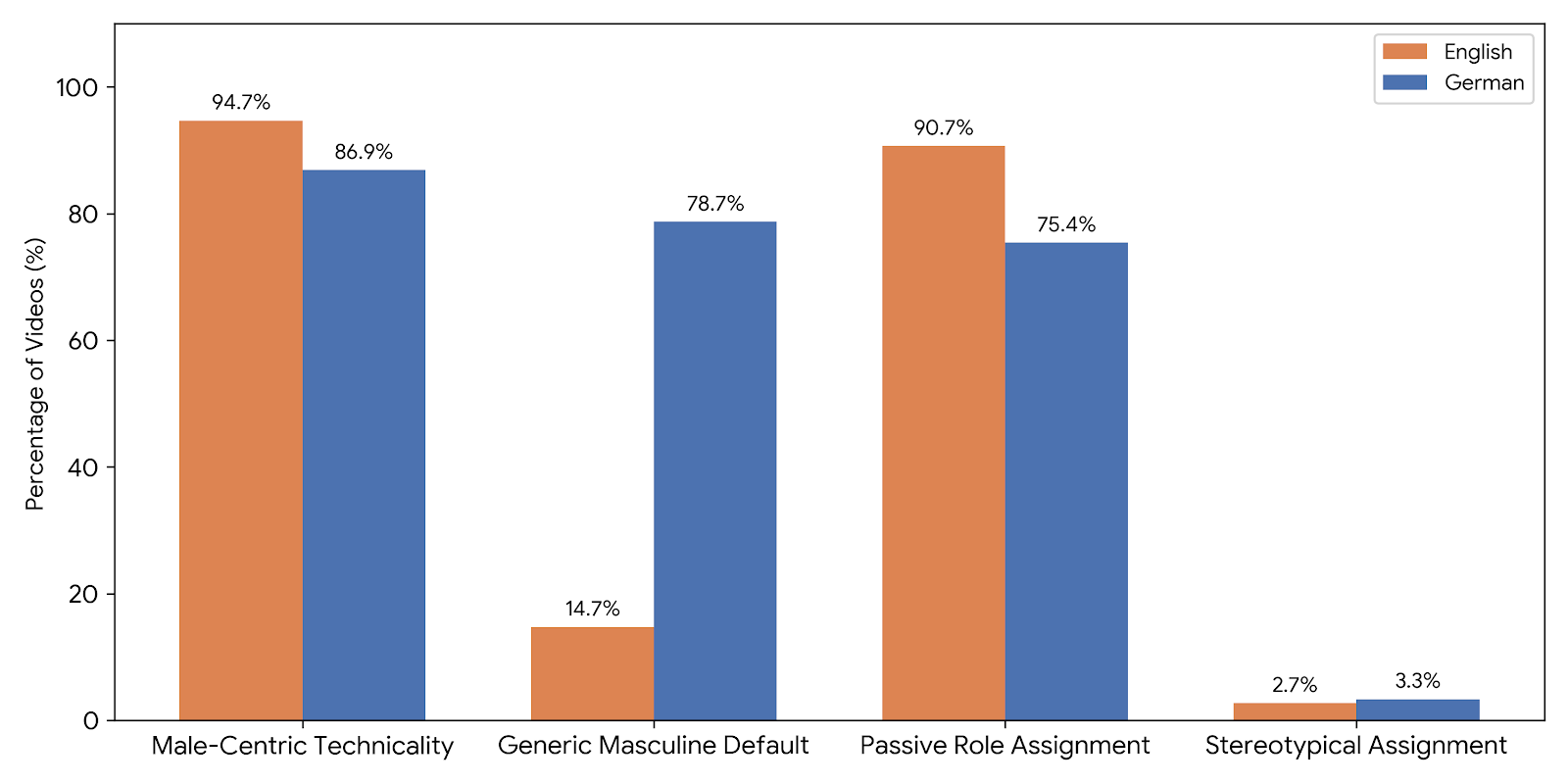} } 
\qquad \subfloat[ Distribution of Gender Representation Index scores. \label{fig:bias-score} ]{ \includegraphics[width=0.44\columnwidth]{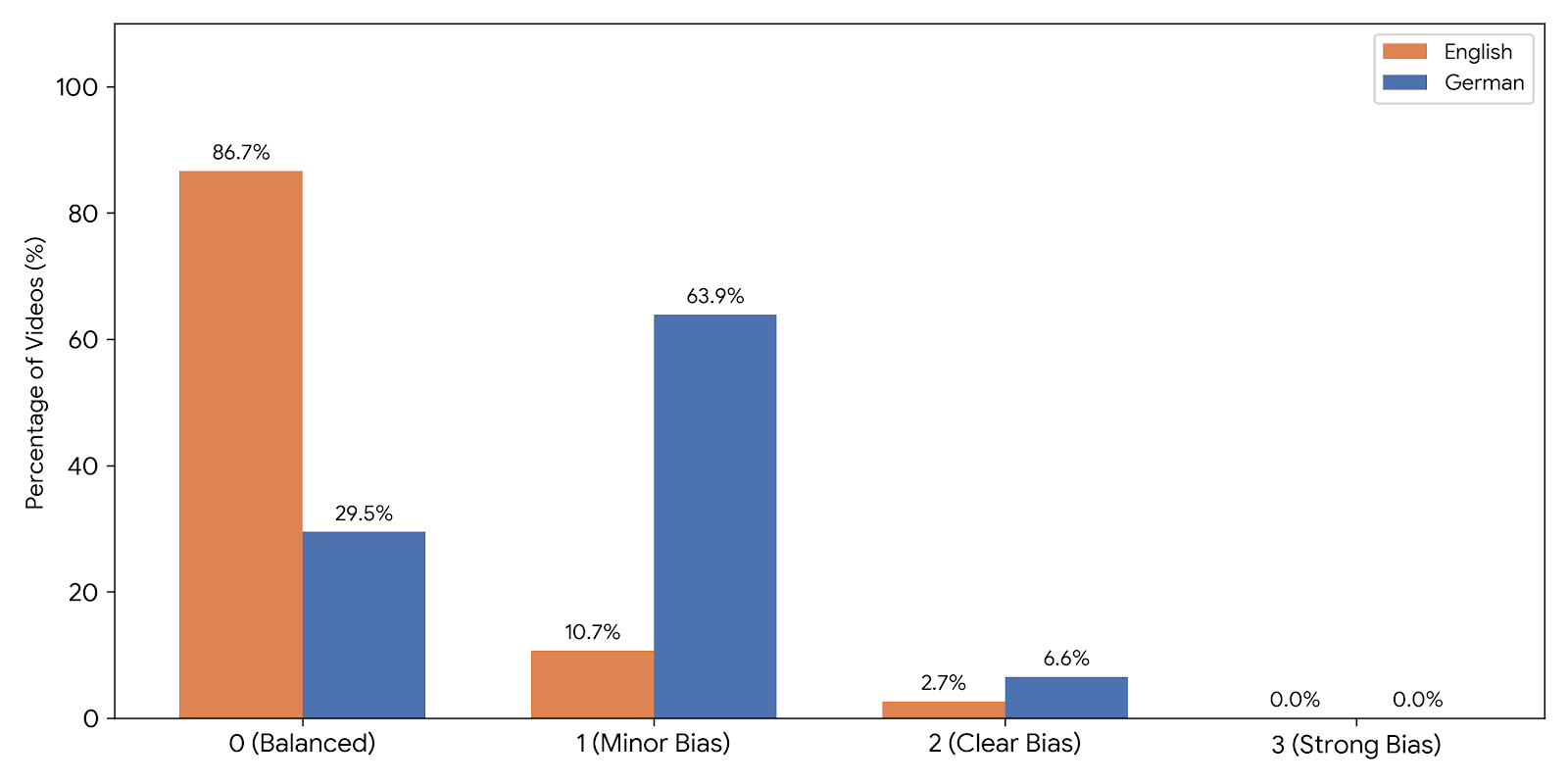} } 
\end{figure}

\Cref{fig:red-flags} illustrates the frequency of bias indicators, which highlight a male-centric and a persistent lack of female agency across both linguistic samples.

In the German sample, the most prominent red flag is the use of the \emph{Generic Masculine as Default}, which occurs in 78.69\% of videos. This is often accompanied by the observation that all technical roles are portrayed as male, resulting in a male-centric technicality (86.89\%). Even when women are represented, they are frequently assigned to \emph{Support or Passive Roles} (75.41\%), such as customers or administrative staff.

The English sample shows that 94.67\% of videos feature technical roles exclusively held by men. However, unlike the German sample, only 14.67\% of English videos trigger the indicator for generic masculine language, due to the standard English use of gender-neutral terms such as \emph{the user} or \emph{they}. The primary indicator for English is \emph{Women in Passive Roles} (90.67\%), indicating that when women do appear, they are almost never the main technical actor.

For both languages, this combination of gendered role assignment creates a glass ceiling within the tutorials, where technical agency is mainly reserved for a male-coded identity.

\subsection{Gender Representation Index} \label{sec:bias-score}
\Cref{fig:bias-score} shows a high-level view of the discursive patterns identified in our sample. The overall index serves as a synthesis of the qualitative integrity of each tutorial, with scores ranging from 0 (Balanced/Neutral) to 3 (Strongly Biased).
The data shows a contrast between the two linguistic samples. German tutorials show a higher average bias score ($0.77$) than English videos ($0.16$). The majority of German content (63.9\%) was categorised as having \emph{Minor Bias} (Score 1), primarily due to the use of the generic masculine. In contrast, 86.7\% of English videos were rated as \emph{Balanced} (Score 0, \Cref{fig:bias-score}). This suggests that while English tutorials often default to gender-neutral phrasing, German content is linguistically predisposed to a masculine-centric discourse.

For videos that reach \emph{Clear Bias} (Score 2), the qualitative data indicate a \emph{double burden} of representation. In the German sample, this score was frequently assigned when masculine pronouns were coupled with specific male-named examples for every active role. For example, in one tutorial (\texttt{pPaVvPJ37YM}), the creator used neutral names to showcase software concepts but defaulted to \emph{der Manager} (manager, male) for the decision-making agent and \emph{die Sekretärin} (secretary, female) for the support role. This reinforces traditional hierarchies, suggesting that even when an effort towards neutrality is made on the surface, the underlying mental models remain gendered.

In the English sample, instances of bias were far less frequent and more subtle. Where a Score 1 was recorded (10.7\%, \Cref{fig:bias-score}), it was almost exclusively due to the repetitive use of \emph{guys} to address the audience or the use of male pronouns for a peer. While some defend the term \emph{guys} as a gender-neutral collective, it remains a point of sociolinguistic debate; its use can signal that the intended or \emph{default} developer is male, and so functioning as a minor but persistent exclusionary cue~\cite{kleinman2021we}.
Notably, neither language sample contained instances of \emph{Strong Bias} (Score 3), indicating that no explicit sexism or exclusion was evident.
\section{Discussion}
Our findings show that while most software engineering education on YouTube appears to be a neutral transfer of technical knowledge, it actually extends a discursive practice that reproduces gendered hierarchies in software engineering. Using the lens of \emph{CDA}, we can interpret these patterns.

\subsection{The Construction of Technical Identity}
We found that as soon as creators introduce a \emph{human example} to explain a concept, e.g., a bank manager, a professor, an architect, or a system administrator, the discourse reverts to the notion that man is in charge. In the German sample, business scenarios consistently assigned technical decision-making to male actors, whereas female actors appeared primarily as consumers or in non-technical support roles.
In \emph{CDA}, agency refers to who is empowered to act within a text~\cite{leeuwen2008}. Our findings show that \emph{High Agency} is a male trait as the discourse constructs the \emph{Architect}, \emph{Developer}, and \emph{Admin} as male identities.

This pattern functions as a \emph{Hidden Curriculum} rooted in traditional gender stereotypes, which is also prevalent in other domains, such as computing education~\cite{dele-ajayi2020} and online retail~\cite{pessach2024}. While the official curriculum teaches software engineering concepts, the underlying semantics associate technical agency and power with masculinity.\footnote{From what we have seen, this masculinity is a cis-heteronormative white-dominated; we would need further research on this topic.} 

When learners consistently encounter male-named characters and male-coded high-agency roles (\Cref{sec:identity}), this might create a (symbolic) barrier to entry for under-represented groups. For female students, the discourse implies that their role in the software lifecycle is that of the \emph{User} or \emph{Customer} (\emph{Low Agency}), while the creative power of engineering remains a male privilege. If female actors do gain agency in these tutorials, it is usually in management rather than technical leadership. This reinforcement may contribute to a sense of \emph{non-belonging}, which might contribute to why under-represented groups either never enter or eventually leave the field~\cite{miller2021}. 

\subsection{Generic Masculine as Naturalised Ideology}
The debate regarding whether the German generic masculine is truly neutral is ongoing \cite{waldendorf2024,brohmer2024}. However, our analysis (\Cref{sec:red-flags}) shows a high correlation between masculine pronouns and male-named characters. From a \emph{CDA} perspective, this indicates a \emph{naturalised ideology}, internalised values that have become an unquestioned standard.
Our data indicates the masculine form is not merely a grammatical placeholder but a reflection of the creator's mental model. If the form were truly neutral, it would appear alongside female names or high-agency female roles more frequently. Instead, technical authority remains male-coded.

\subsection{The Male Mental Model}
We can put this in perspective by directly interpreting three different English videos covering \emph{Agile Software Engineering Practices} from the results.

\begin{figure}
    \centering
    \includegraphics[width=0.4\linewidth]{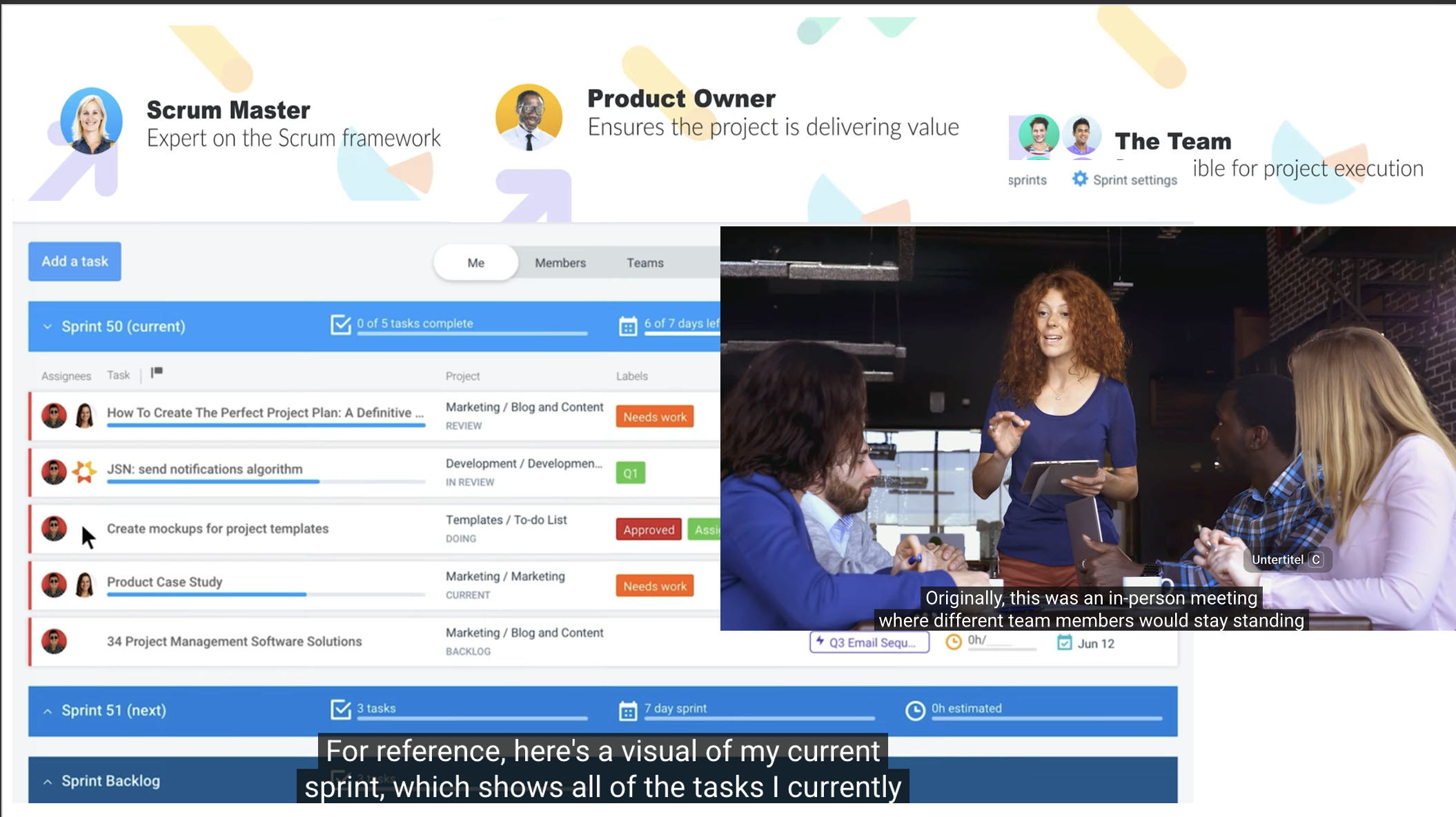}
                \caption{Balanced gender representation in an agile context ({\tiny ZRMwR5NnHXY}).}
    \label{fig:positive}
\end{figure}

\emph{The guy is the leader (\texttt{RQ--Sells\_g}):} The male narrator uses neutral \emph{they/them} for the general team members. However, when referring to the \emph{leader} of the team and process (Product Owner), he switches to \emph{guy} and \emph{he}. This subtle shift makes it clear who is \emph{naturally} expected to lead.

 \emph{Balanced Representation (\texttt{ZRMwR5NnHXY}):} The male narrator avoids gendered language when referring to the team and their roles. However, the accompanying images show a diverse team in terms of gender and ethnicity. This video demonstrates that team composition can vary (\Cref{fig:positive}).

 \emph{The rare female High-Agency (\texttt{4pZt9isVymQ}):} Notably, the \emph{only} video in our entire subsample that features a technical high-agency role without any male involvement was produced by a Black woman in tech, using herself as the primary example.
The fact that women often appear as \emph{the only one} (e.g., combinations of \texttt{0m1f0n}) is symptomatic of tokenism. In these tutorials, one woman might be often seen as \emph{enough} to represent diversity, which does not reflect a truly balanced professional environment. Since YouTube uses popularity-based algorithms, if these biased patterns remain popular, the platform will continue to show this bias to future generations of developers~\cite{kanetaki2022}.

\subsection{Implications and Future Work}
Our findings suggest that the software engineering community should be far more aware and precise in its use of language and examples. This should not be dismissed as a \emph{soft} issue or a purely humanities-based concern; rather, it is a matter of technical and professional accuracy. 
This shift in practice can be seen as similar to the debate and move to replace exclusionary technical language, such as the \emph{master-slave} analogy~\cite{eglash2007}. Just as we refactored our technical vocabulary to better reflect modern values, we must address the \emph{dude culture} that persists~\cite{wang2024,miller2021}. 
From our perspective, it seems feasible to improve the gender representation in the tutorials by using diverse naming for human examples, neutral pronouns, and being aware of balanced role assignments. 

Since our work identified a male-centric hidden curriculum, a next step would be to examine the effects on students of gender-neutral or female-led tutorials to assess shifts in their self-efficacy. Those experiments would test \emph{motivational role model theory} and \emph{stereotype threat}. To validate our \emph{CDA} finding, future research should expand the analysis to other languages, e.g., Romance languages such as French, to compare masculine defaults, and to genderless languages to test whether low gender representation persists when the language is neutral. 
We focused only on a binary gender category; however, it is important to include non-binary and non-cis learners and content creators, and connect with prior research on transgender developers~\cite{ford2019}. This is especially interesting with the background of they/them use offers more challenges for non-binary individuals~\cite {saguy2022little}. \emph{Intersectionality} is more important than ever in the current discourse, so we would like to investigate how ethnicity and disability intersect with gender, e.g. indigenous software practitioners~\cite{sanchez-gordon2025}.

\section{Conclusions}
This study presents the first empirical evidence of gender representation in software engineering educational online material. 
Our analysis shows that, while English discourse has trended toward linguistic neutrality, both languages continue to struggle to represent women in situational contexts, particularly in high-agency technical roles, as the technical examples remain overwhelmingly male-centric.
If we want these tutorials to reflect the diversity we hope to see in the software industry, we need to acknowledge these discursive patterns as a first step towards making a more thoughtful design of examples. 
We do not argue for an artificial or radical overhaul, but rather for a move toward representational accuracy. By including more female names, giving women more active roles in coding examples, and using a wider range of settings, creators can break the hidden message that being in the software engineering community is a male trait. We hope to encourage educators and researchers to address this in future work to contribute to a learning environment where everyone is visible.

%

\bibliographystyle{splncs04}
\bibliography{references}

\end{document}